\documentclass[preprint,12pt, sort&compress]{elsarticle}

\usepackage{amssymb}
\usepackage{color}
\usepackage{soul}
\usepackage{chemformula} 
\usepackage[T1]{fontenc} 
\usepackage{graphicx}
\usepackage{subcaption}
\usepackage{cleveref}
\usepackage{color}
\usepackage{ulem}
\usepackage{derivative}
\usepackage{adjustbox}
\usepackage{makecell}
\usepackage{soul}
\usepackage{booktabs}
\usepackage{bm}
\usepackage{lmodern}
\usepackage{fix-cm}

\begin{document}

\begin{frontmatter}

\title{Underwater bubble transport on superhydrophobic cylindrical rod}

\affiliation[inst1]{organization={Multiscale Multiphysics Group (MMG), Department of Mechanical Engineering},
            addressline={Indian Institute of Technology Madras}, 
            city={Chennai},
            postcode={600036}, 
            country={India}}
\affiliation[inst2]{organization={Department of Biological and Environmental Engineering},
            addressline={Cornell University}, 
            city={Ithaca, New York},
            postcode={14853}, 
            country={USA}}
\author[inst1]{Rajalingam A}
\author[inst2]{Sunghwan Jung}
\author[inst1]{Pallab Sinha Mahapatra}
\ead{pallab@iitm.ac.in}


\begin{abstract}

\textbf{Hypothesis:} The transport of bubbles along a curved superhydrophobic surface is not governed by buoyancy alone, but also by the interaction between the geometric confinement, the capillary and contact-line resistance, and the hydrodynamic resistance. We hypothesize that the ratio of rod-bubble diameter controls bubble attachment, transport, and detachment, and that the wake produced by one bubble can also influence the transport state of a subsequent bubble.

\textbf{Experiments:} Air bubbles of various diameters were transported along submerged cylindrical rods with different diameters and inclination angles. We systematically investigated the capillary number ($Ca$) and Bond number ($Bo$) of the transported bubble, as well as the size of the rod relative to the bubble. An analytical force-balance model was developed to predict transport velocity, accounting for buoyancy, hydrodynamic drag, and capillary resistance. The transition from transport to detachment was identified. The experiments were carried out sequentially and supplemented by axisymmetric numerical simulations to investigate wake-induced interactions between bubbles.

\textbf{Findings:} Bubble transport is strongly influenced by rod inclination and the bubble-to-rod size ratio, demonstrating that curvature-induced confinement alters the balance between driving and resistive forces. The analytical model captured the experimental trends with deviations of less than 10\% in most cases. The bubble detaches from the inclined rod at a critical Bond number that depends on rod diameter, facilitating the development of a transport regime map. The study also found that the motion of successive bubbles is strongly coupled. When a following bubble enters the wake of a preceding bubble, it speeds up and attains a higher capillary number than the leading bubble. Both experiments and numerical simulations consistently reproduce this wake-mediated acceleration, which shows that bubble transport is governed by both the force balance on individual bubbles and the hydrodynamic interactions between adjacent bubbles. These findings form the basis for controlling bubble transport, detachment, and collective motion on superhydrophobic interfaces.

\end{abstract}




\end{frontmatter}

\section{Introduction}

The controlled transport of gas bubbles in an aqueous medium has received significant attention due to its role in multiphase transport, interfacial engineering, and energy-related applications. The motion of the bubble on the structured solid surfaces affects mass transfer, hydrodynamic resistance, and phase-separation efficiency in various systems, including electrochemical reactors, wastewater treatment units, flotation processes, and microfluidic devices \cite{betz2013boiling, van2018gas, stanic2021bubble, yu2023bubble,chowdhury2022wettability}. Efficient removal of gas bubbles is important in electrolysis and fuel-cell systems. Bubbles that form on the electrode surface can cover active sites and increase electrical resistance, thereby reducing system performance \cite{andaveh2022superaerophobic, he2026advances}. Prompt bubble detachment and directed bubble transport can significantly enhance operational efficiency \cite{he2026advances,li2016single}. The detachment of the bubble from the liquid-liquid interface is also another challenging problem for the chemical industries \cite{chowdhury2022wettability}.

Surface modification by altering wettabilities \cite{josyula2024fundamentals} and appropriately patterning \cite{sinha2022patterning} helps manipulate fluids. Superhydrophobic surfaces offer an attractive way to passively manipulate bubbles due to their low liquid adhesion, trapped air layers, and minimized hydrodynamic drag \cite{ling2011increased, costantini2018drag, fomicheva2026advanced}. In underwater environments, these surfaces exhibit a strong attraction to air, allowing bubbles to attach, move, and merge along specific paths with ease \cite{yong2019substrate,gurera2021movement}. These qualities have inspired the development of the surface for guiding gas designs for applications such as gas collection, lab-on-a-chip transport, flotation-assisted oil cleanup, and controlled aeration \cite{arscott2013wetting, yuxin2026superhydrophobic}. Structured surfaces guide bubble motion through capillary and pressure-driven mechanisms \cite{zhang2016particle,gurera2020contact}. They have shown promise in improving energy-efficient gas-liquid transport \cite{van2018gas, jeung2020underwater}.

Previous studies have demonstrated that underwater directional bubble transport occurs in many superhydrophobic structures, particularly on patterned surfaces with a wettability gradient and a wedge-shaped surface \cite{ahmed2026rise,chen2018self,zhu2026directional}. Patterned slippery surfaces soaked with lubricating oil have been shown to enhance bubble collection and directed movement by reducing interfacial resistance \cite{li2019patterned,tang2018bioinspired}. The impact of lubricant viscosity, tilt angle, and surface features on bubble velocity has also been studied, which indicates that the lower viscous resistance and higher angles allow faster transport \cite{tang2018bioinspired}. Other studies have looked at bubble manipulation using patterned superhydrophobic substrates \cite {zheng2024demand}, helical structures \cite{yu2016superhydrophobic}, rails \cite{zhu2020spontaneous}, and sloped hydrophobic surfaces \cite{ali2017bubble}. These studies have shown that surface shape significantly affects bubble adhesion, migration speed, and transport stability. Although several shapes were used in existing studies, as noted above, conical superhydrophobic structures have received particular attention because they generate a pressure gradient due to curvature changes along the cone's axis. This pressure gradient causes bubbles to move directionally, even against buoyancy \cite{yu2016spontaneous,xue2016superhydrophobic,li2019bubble}. Earlier research indicated that increasing the angle of the cone increases the velocity of the bubble due to a stronger driving force \cite{xue2016superhydrophobic}. The balance of buoyancy, drag, pressure, and contact-angle hysteresis forces has also been analyzed to understand bubble behavior on conical surfaces \cite{xue2018reliable}. However, the overall distance traveled, and the associated velocities are generally low on the conical surface \cite{ma2018directional}. Recently, confined transport between paired superhydrophobic structures has been shown to enhance transport efficiency by reducing bubble deformation \cite{du2025bubbles, gao2025influencing}.

Droplet transport along cylindrical fibers and wires has been studied extensively for fog harvesting \cite{moncuquet2022collecting}. Previous research shows that droplet motion depends on capillary forces, wettability, viscous dissipation, and the shape of the substrate \cite{lee2022multiple,leonard2023droplets,schwarzwalder2023experimental,wang2025spontaneous,jiang2022coalescence,mumm2009easy,wang2005experimental}. In these systems, the phase being transported is liquid, while the surrounding phase is gas. In contrast, this study looks at the opposite. This switch between the dispersed and continuous phases creates different conditions for the surface and fluid dynamics. This includes the effect of the surrounding liquid, viscous resistance, pressure distribution, and the wake formed behind the moving bubble. Specifically, unlike a liquid droplet on a fiber, the gas bubble creates a liquid flow field and a wake that can affect other subsequent bubbles. Thus, the behavior of bubble transport along a rod cannot be directly linked; therefore, a detailed study is warranted.

Recent studies on superhydrophobic wires have shown sliding bubble transport and bubble-mediated cargo movement, indicating that both the bubble–wire configuration and the size of the bubble can affect the rising path and the transport velocity \cite{yu2016superhydrophobic, zhang2020efficient, zhang2025spontaneous}. Although there have been recent improvements in the directional transport of bubbles on structured superhydrophobic surfaces, bubble transport on cylindrical substrates has not been studied adequately.  The combined influence of bubble size, cylindrical rod size, and rod inclination angle on bubble transport and detachment remains poorly understood. Even more importantly, the movement of one bubble after another along a submerged superhydrophobic structure, and the hydrodynamic interactions between adjacent bubbles, have received limited attention. In this study, we developed a systematic experimental, analytical, and numerical approach to clarify the coupled transport mechanisms on submerged superhydrophobic cylindrical rods. First, the effects of bubble diameter, rod diameter, and rod inclination are determined, and the transition from stable bubble transport to detachment is investigated in detail. A force-balance model that accounts for buoyancy, hydrodynamic drag, and capillary hysteresis is developed to predict bubble transport velocity and is validated against experimental measurements. Beyond the case of a single isolated bubble, the study shows that a preceding bubble can enhance the transport of a subsequent bubble, as demonstrated through a combination of experiments and numerical simulations. It reveals the roles of the induced wake flow, local liquid circulation, and interfacial conditions in this cooperative transport. These results provide a mechanistic understanding of bubble transport on cylindrical superhydrophobic interfaces and lay down design principles for passive bubble manipulation in electrochemical systems, gas–liquid separation, flotation, and microfluidics.

\begin{figure*}[ht!]
\centering
\includegraphics[width=0.98\textwidth]{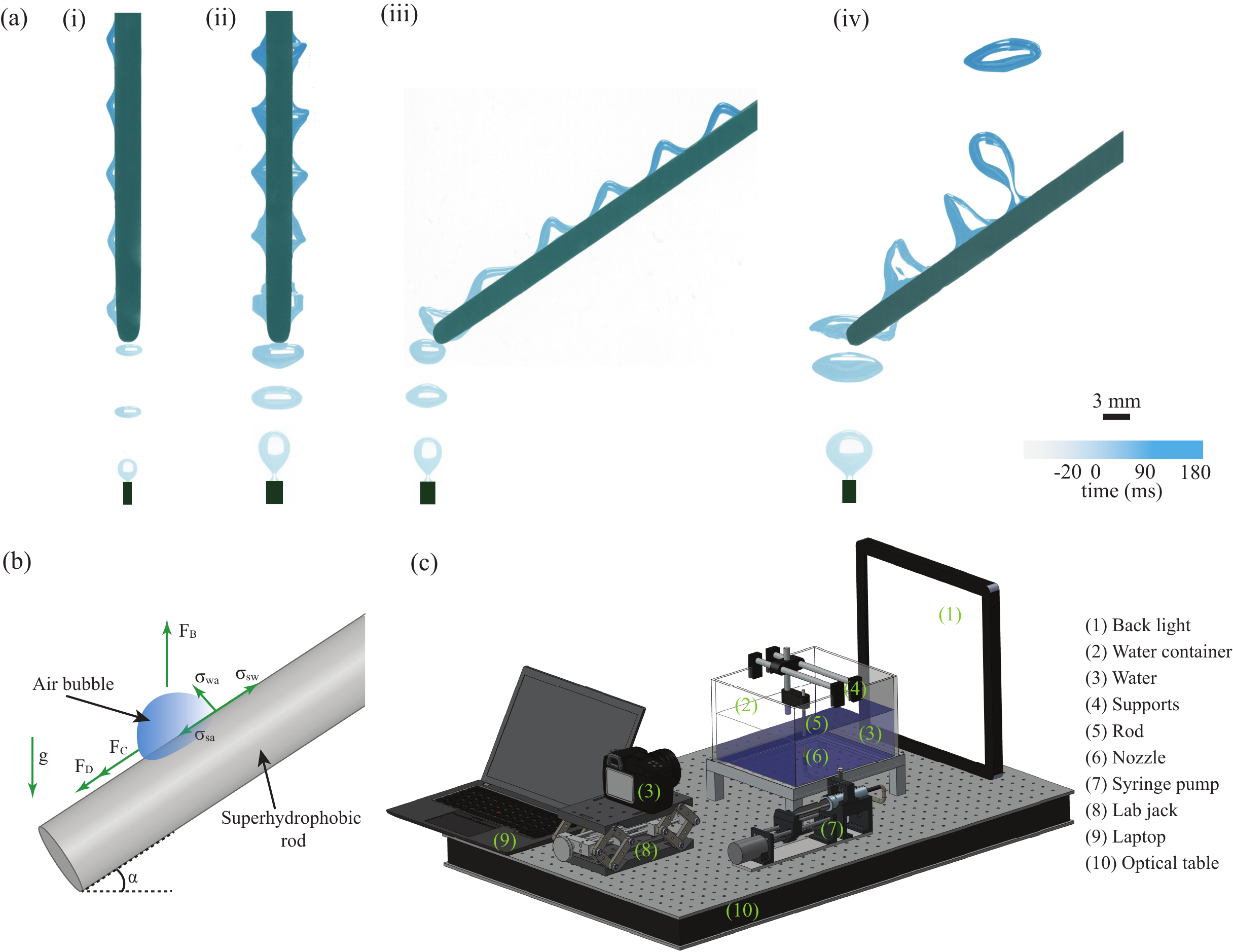}
\caption{Bubble transport on rods, forces acting on bubbles during transport, and the experimental setup. (a) Bubble transport on a 3 mm diameter rod: (i) vertically oriented rod with an equivalent bubble diameter of 2.2 mm; (ii) vertically oriented rod with an equivalent bubble diameter of 3.8 mm; (iii) transport of a 2.8 mm equivalent diameter bubble on a rod inclined at $30^\circ$; (iv) transport of a 6 mm equivalent diameter bubble on a rod inclined at $30^\circ$. (b) Schematic representation of the forces acting on a bubble during transport along a circular rod. (c) Schematic of the experimental setup for studying bubble transport.}
\label{geo}
\end{figure*}

\section{Materials and methods}
\label{sec:E_M}

\subsection{Surface preparation}

An aluminum (6061 alloy) rod was used as the substrate material and shaped into cylinders with diameters of $3, 6$, and $8$ mm. The lower end of each rod was curved to reduce unwanted air trapping during immersion. The original aluminum surface is naturally hydrophilic. Therefore, it was chemically modified as a superhydrophobic surface. The rods were first cleaned in an ultrasonic bath using acetone (99\% pure, purchased from SRL chemicals) and deionized water (produced by Lablink water purification system, Model – XTRA Pure RO) to remove contaminants. Next, a 3 M hydrochloric acid (HCl, approximately 37 wt\%, ACS reagent grade; Emparta) solution was used to etch the cleaned samples, creating microscale roughness. After etching, the samples were rinsed and heated with deionized water at 100\textdegree C for one hour, forming a superhydrophilic oxide layer. To achieve superhydrophobicity, the silicon dioxide nanoparticles (Glaco, Soft99 Corporation) were deposited onto the etched surface by spraying. The combination of etching and nanoparticle application created a micro/nanostructure that effectively traps air and repels water. The goniometer (Dataphysics OCA 11) was used to characterize the surface wettability. The final surfaces had a static water contact angle of 156\textdegree $\pm$ 4\textdegree, confirming their superhydrophobic characteristics (See S1).

\subsubsection{Experimental Procedure}

Figure~\ref{geo}c shows the bubble transport behavior, the forces acting on a bubble moving along the rod, and the experimental setup fabricated in-house for the present study. The experimental setup consists of a water container, syringe pump (New Era Pump Systems-300), stainless-steel nozzle, assembly of superhydrophobic rod, high-speed camera (Phantom, VEO 340), LED backlight, and mechanical supports. The high-speed camera and backlight were aligned with the test section in between to capture shadowgraph of bubble dynamics. The container was filled with clean distilled water at room temperature. Air bubbles were injected using a syringe pump connected to a submerged nozzle through a flexible tube. The nozzle was placed 15 mm below the rod so that the bubbles could reach nearly terminal velocity before interacting with the rod surface. 

During each experiment, air bubbles were released into the water and allowed to rise toward the superhydrophobic rod. After reaching the rod, the bubble moved along its length. A high-speed camera was used to capture the bubble motion (with 2,000 fps). The recorded images were analyzed using ImageJ and MATLAB. The video was first extracted into individual frames, and the frames were converted to grayscale. The image contrast was then adjusted, followed by binary thresholding for distinguishing the bubble from the surrounding water. The leading edge of the bubble was tracked in each frame, and its position was recorded with time. The bubble velocity was obtained from the slope of the displacement–time plot using a linear fit.

The experiments were conducted for equivalent bubble diameters of $2.2$, $2.8$, $3.8$, $4.8$, and $6$ mm with a measurement uncertainty of about $\pm$0.1 mm. The equivalent diameter was measured immediately after the bubble detached from the nozzle by fitting a circle to the bubble's contour at its closest spherical shape. Each experiment was repeated three to five times to ensure reliability. To verify the experimental methods, additional trials were performed with freely rising bubbles in water, and the measured bubble velocities were compared with previously reported values (See S3).

\begin{figure}[ht!]
\centering
\includegraphics[width=1\textwidth]{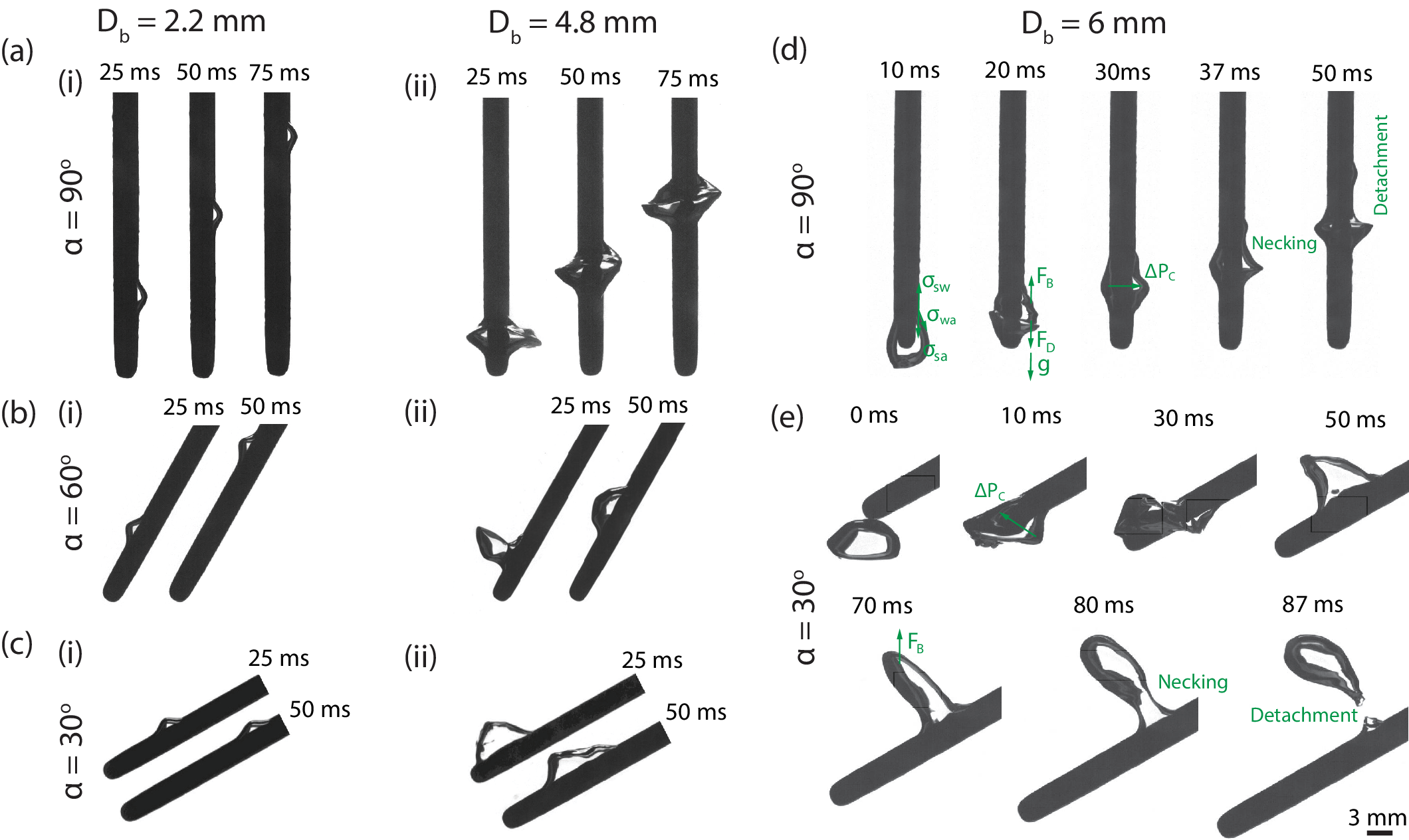}
\caption{Transport of a bubble on a superhydrophobic rod for various bubble diameters and tilt angles: (a) bubble transport with the rod oriented vertically (b) bubble transport at a tilt angle of $ 60^\circ$ (c) bubble transport at a tilt angle of $30^\circ$ (d) bubble transport detachment when the rod is oriented vertically (e) bubble detachment during transport for a rod oriented at a tilt angle of $30^\circ$.}
\label{exp2}
\end{figure}

\section{Results and discussions}

which were then To generalize the transport dynamics of bubbles along superhydrophobic rods, the motion was characterized using the dimensionless Bond number ($Bo$), which represents the ratio of gravitational buoyancy to surface tension forces and is defined as $Bo = \frac{\Delta \rho g D_b^{2}}{\sigma}$. Here, $\Delta \rho$ is the density difference between the phases, $g$ is the gravitational acceleration, $D_b$ is the equivalent bubble diameter, and $\sigma$ is the surface tension. The transport velocity is expressed through the Capillary number $Ca = \frac{\mu U}{\sigma}$, where $\mu$ is the dynamic viscosity of the liquid and $U$ is the terminal velocity of the bubble. 
 
\subsection{Bubble Transport behavior on Superhydrophobic rods}
 \label{subsecforces}
 Figure \ref{exp2} illustrates the transport dynamics of air bubbles with varying diameters (\(D_{b}\)) on a superhydrophobic rod of 3 mm diameter (\(D_{p}\)) at various angles of inclination. Based on the relative size of bubbles and rods, three different regimes are identified: (i) $D_b \ll D_p$, (ii) $D_b \approx D_p$, and (iii) $D_b \gg D_p$. In the first regime, $D_b \ll D_p$, the bubble makes the contact only a small part of the rod circumference. The bubble remains nearly spherical with a relatively flat contact region and moves steadily along the rod with only small changes in its interface, as shown in Fig.~\ref{exp2}a-i. In the second regime, $D_b \approx D_p$, the bubble covers a larger portion of the rod circumference. Due to the space constraint around the rod, the three-phase contact line becomes more curved. In the third regime (\(D_b \gg D_p\), see Fig.~\ref{exp2}a-ii), the bubble completely envelopes the rod, causing severe interfacial deformation and shape instability driven by the interplay of buoyancy, inertia, hydrodynamic drag, and surface tension. Figures \ref{exp2}b and c depict transport at tilt angles of $60^\circ$ and $30^\circ$, respectively. In both cases, the bubbles preferentially migrate along the upper ridge of the rod, driven by the synergistic effects of buoyancy and curvature-induced capillary forces.

Figure~\ref{exp2}d and e show the transport behavior of a bubble with a diameter of $D_b = 6$ mm on a vertically and an inclined oriented rod ($30^\circ$), respectively, where the distinct bubble detachment behavior has been observed. To achieve a holistic understanding of the bubble transport phenomena and detachment on a superhydrophobic rod, the governing forces have been investigated (see Fig.~\ref{geo}b) and a scaling analysis of the transport behavior. The motion of a bubble along an underwater superhydrophobic rod is determined by the dynamic equilibrium between the driving and resistive forces. The behavior of bubble transport on a superhydrophobic rod immersed in water results from the balance of the driving and retarding forces. As shown in Eq.~(\ref{force_eqn}), tangential force acts on the bubble, where the buoyancy force ($F_B$) drives the bubble motion when the drag force ($F_D$) and the force due to contact angle hysteresis ($F_C$) act in the opposite direction. 

\begin{equation}
    \rho \Omega \left( C \frac{dU}{dt} \right)
= F_{\text{B}} - F_{\text{D}} - F_{\text{C}}
\label{force_eqn}
\end{equation}

At the terminal velocity of the bubble, $\frac{dU}{dt} = 0$. The buoyancy force is expressed as $F_B \approx\Delta\rho \, V_b \, g$. 
It indicates a strong increase with bubble size as the buoyancy force scales with the cube of the equivalent diameter of the bubble ($F_B \propto D_b^3$) for a spherical bubble that aids in bubble motion. 

Two distinct capillary interactions must be considered at the three-phase contact line. The first is that the static capillary force acts perpendicular to the solid surface, resulting from the liquid-gas surface tension ($F_{C}\sim\sigma$). It helps retain the bubble's attachment to the rod, though it does not affect any component of the bubble's lateral transport vector. In addition, the angles of advance and recession may differ due to surface roughness, defects, or chemical heterogeneity. The asymmetric wetting front generates a capillary resistance force ($F_{C}$). This force can approximately expressed as as  $F_{C} \approx \sigma \, w \, (\cos\,\theta_R - \cos\,\theta_A)$, where $ \sigma$, $w$, $\theta_R$, and $\theta_A$ are the surface tension of the liquid-gas, the characteristic dimension, the receding contact angle, and the advancing contact angle, respectively. Although capillary pinning is very low on the superhydrophobic surface, it defines the shape of the bubble base and contributes to contact line deformation, especially when the base motion is irregular. When the bubble is modeled as a spherical cap residing on the cylindrical rod with a high apparent contact angle ($\theta = \pi$), the hysteric force can be scaled to $F_C \sim w \sigma (\theta_R^2-\theta_e^2)$, where the characteristic length $w$ scales with the radius of the rod ($w \sim R_p$). Using the Hoffman-de-Gennes equation \cite{kim2017capillary}, $\theta_A(\theta_R^2-\theta_e^2) = \frac{1}{0.013}Ca$. Here, $Ca \sim \frac{\mu U}{\sigma}$. Therefore, $F_C \sim \frac{\mu U R_p}{\theta_A}$.

At high Reynolds numbers (Re$\gg$ 1), hydrodynamic drag is another significant resistive force as expressed in $F_D \sim \frac{1}{2}\rho_lU^2A_p $, where $\rho_l$ is the density of water, $A_p$ is the projected frontal area of the bubble, and $U$ is the velocity of the bubble. As the bubble is assumed to form a spherical cap around the cylindrical rod, its projected frontal area can be expressed as $A_p=L\times H=\frac{4}{3\pi}\frac{R_b^4}{D_p^2}$. Here, $L$ and $H$ denote the length and height of the bubble, respectively, on the superhydrophobic rod. However, the presence of the rod significantly modifies the bubble's projected area and shape, as well as the local flow field in the surrounding liquid. Additionally, the bubble tends to move along the entrapped air layer (plastron layer) on the rod's surface roughness, thereby reducing resistance by inducing slip \cite{fomicheva2026advanced}. As a result, the presence of the rod increases the bubble velocity, resulting in a higher velocity than that of a freely rising bubble.

By substituting the explicit scaling expressions for buoyancy, dynamic capillary pinning, and hydrodynamic drag into the terminal force balance Eq.~(\ref{force_eqn}), we arrive at the following governing quadratic equation for steady-state bubble transport as
\begin{equation}
     \frac{1}{2}\rho_l\frac{4}{3\pi}\frac{R_b^4}{D_p^2} U^2 + \frac{\mu R_p}{\theta_A} U - \frac{4}{3\pi}\rho_l g R_b^3 \sin \alpha = 0
\label{force_qua}
\end{equation}

The solution of Eq.~(\ref{force_qua}) yields the theoretical prediction for the steady-state bubble transport velocity as

\begin{equation}
 U^{Theor} =
\frac{
-\frac{\mu R_p}{\theta_A}
+
\sqrt{
\left(\frac{\mu R_p}{\theta_A}\right)^2
+
\frac{32}{9\pi^2}
\frac{R_b^4}{D_p^2}
\rho_l^2 gR_b^3\sin\alpha
}
}{
\frac{4\rho_l R_b^4}{3\pi D_p^2}
}
\label{force_soln}
\end{equation}
A dimensionless proportional parameter will be introduced to account for the velocity calculation in the presence of non-idealities, such as geometric deviations from the spherical-cap assumption and local variations in plastron stability. Further, the capillary number will be calculated using the calculated velocity as $Ca^{Theor} = \frac{\mu U^{Theor}}{\sigma}$. This analytical hypothesis will be systematically validated against experimental kinematics in the following section.

\begin{figure}[ht!]
\centering
\includegraphics[width=0.95\textwidth]{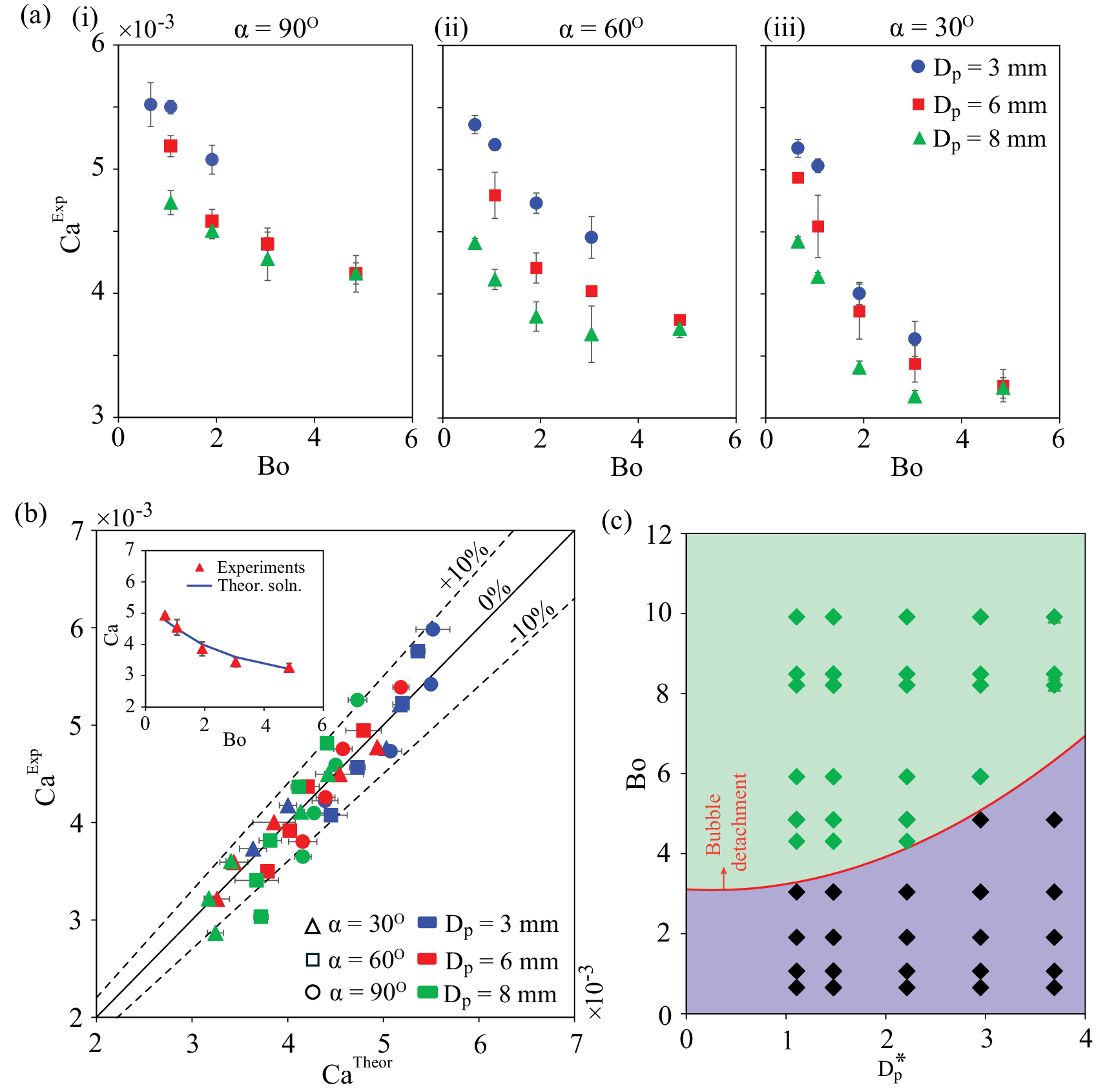}
\caption{Variation of capillary number ($Ca$), comparison between experimental and theoretical $Ca$, and regime map: (a) variation of capillary number ($Ca$) as a function of Bond number ($Bo$) for various rod diameters ($D_p = 3, 6$ and $8$ mm) and tilt angles of the rod: (i) $\alpha = 30^\circ$, (ii) $60^\circ$, (iii) and $90^\circ$. (b) Comparison of the theoretically predicted capillary number ($Ca^{Theor}$) with the experimental capillary number ($Ca^{Exp}$) (c) regime map showing bubble transport along the rod and bubble detachment from the rod.}
\label{fig:Bo_ca}
\end{figure}

As highlighted previously, Figs.~\ref{exp2}d and e illustrates the bubble (\(D_b = 6\) mm) transport and detachment on rods. On a vertical rod (see Fig. ~\ref{exp2}d), the bubble spreads across the superhydrophobic surface wraps around most of the rod's perimeter due to capillary adhesion. However, the asymmetric wrapping induces variations in interfacial curvature and creates a strong Laplace pressure imbalance. Simultaneously, buoyancy pulls up the bubble upward while hydrodynamic drag holds it back on the surface of the rod. This fierce competition of forces induces rapid interfacial oscillations and localized necking, which quickly terminates in bubble pinch-off. A similar mechanism is observed for inclined rods (Fig.~\ref{exp2}e). The bubble first spreads over the rod and then moves toward the upper ridge due to the combined effects of buoyancy and the curvature-induced pressure gradient. This upward stretching caused by buoyancy reduces the solid–liquid contact area and, consequently, the capillary retention force. Therefore, the transition from bubble transport to detachment is governed by the balance of buoyancy and capillary forces. Buoyancy is characterized by the Bond number ($Bo \sim \rho g D_b^2/\sigma$), whereas the capillary retention force scales with the characteristic length as $F_{C}\sim w$. 
Since $w$ is determined by the rod geometry, we use the normalized rod diameter, $D_p^*=D_p/\sqrt{\sigma/(\rho g)}$, to account for the effect of rod size. Based on these two parameters, we construct a regime map of $Bo$ versus $D_p^*$ (Fig.~ \ref{fig:Bo_ca}c). The map shows that increasing $Bo$ promotes bubble detachment, while increasing $D_p^*$ favors bubble retention.

\subsection{Effect of bubble and rod diameter}

The transport dynamics of bubbles along superhydrophobic rods are characterized using the dimensionless Bond number ($Bo$) and the Capillary number ($Ca$). Figure~\ref{fig:Bo_ca}a shows the variation of $Ca$ with $Bo$ for different diameters of the rod ($D_p = 3, 6,$ and $8$ mm) and inclination angles ($\theta = 30^\circ, 60^\circ$, and $90^\circ$). For rods with diameters of $3$ mm and $6$ mm, $Ca$ is inversely related to $Bo$. Although an increase in $Bo$ represents an increase in buoyancy drive, it also causes the area of the bubble's footprint on the rod surface to increase. In this case, the resistive capillary forces determined by the scale of the triple-phase contact line, the deformation of the bubble interface, and the drag take precedence over the volumetric buoyancy force. As a result, the terminal velocity of the bubble decreases, as is shown by the decreasing $Ca$ values with increasing bubble size. 

The role of substrate geometry is also evident in the change in $Ca$ with rod diameter; for a given $Bo$, $Ca$ decreases as rod diameter increases. This is due to the larger wetting arc of the larger rods, which amplifies the effect of interfacial resistance, thereby increasing the total resistive force exerted on the bubble. Moreover, $Ca$ has a strong dependence on the inclination of the rod; as the tilt angle decreases from the vertical ($90^\circ$), the effective axial component of the buoyancy ($F_b \sin\alpha$) is reduced. This decrease in the net driving force leads to the observed reduction in bubble velocity, further demonstrating that the transport process is governed by the balance between gravitational drive and geometric confinement.

It is interesting that the $8$ mm rod shows a non-monotonic behavior at inclination angles of $30^\circ$ and $60^\circ$. At these angles, $Ca$ first decreases until a critical Bond number of about $Bo \approx 3.04$ is attained, after which $Ca$ starts to increase as the Bond number increases. This change indicates a transition between the two dominant force regimes; at lower $Bo$ values, the large wetting arc of the $8$ mm rod gives enough pinning to prevent motion. However, when the bubble size exceeds this critical value, the buoyant force eventually overcomes the pinned contact-line resistance, and the velocity increases. Moreover, for a given $Bo$, $Ca$ tends to decrease as the rod diameter increases and the tilt angle decreases. This is because the axial component of the buoyancy force ($F_b \sin\alpha$) is reduced, thus lowering the net driving force for transport.

The experimental hypothesis discussed above regarding bubble transport behavior on the superhydrophobic rod has been compared with the theoretical model derived in Eq.~\ref{force_soln}. In the equation, the proportionality constant \(C\) was fine-tuned using MATLAB regression and was found to range from 1.2 to 1.55. The advancing contact angle is measured from time-lapse experimental images, and the average values are used to calculate the bubble velocity. Further, the capillary number is calculated using the calculated velocity as $Ca^{Theor} = \frac{\mu U^{Theor}}{\sigma}$. They are compared with the experimental capillary number ($Ca^{Exp} = \frac{\mu U^{Exp}}{\sigma}$) and are shown in Fig. ~\ref{fig:Bo_ca}b. The figure shows that the deviation between the experimental and analytical values is less than $10\%$ in most cases. The subplot shows that the theoretical model accurately predicts the trend of the experimental results.

\begin{figure}[ht!]
\centering
\includegraphics[width=0.97\textwidth]{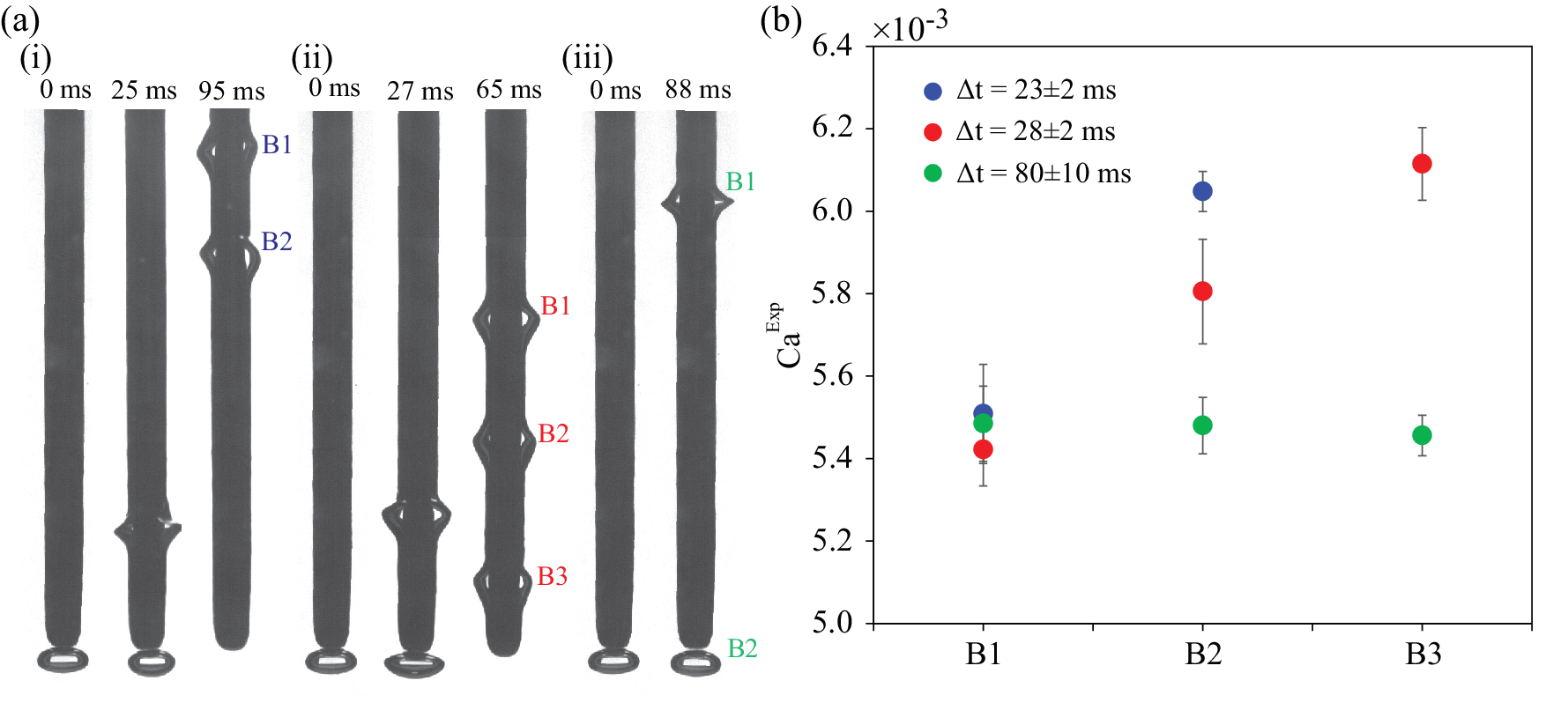}
\caption{Effect of preceding bubble in the velocity of subsequent bubble (a) Multiple bubble transport behavior on the rod with the diameter of 3 mm oriented vertically (b) Capillary number of bubbles transported in series.}
\label{bubble_chain}
\end{figure}

\subsection{Effect of preceding bubble on the motion of subsequent bubble}

In order to study the cooperative transport behavior in a continuous bubble chain, the effect of preceding bubbles on the transport velocity of the subsequent bubbles was systematically investigated. The experiments were conducted over three different time intervals: \(\Delta T = 23 \pm 2\) ms, \(28 \pm 2\) ms, and \(80 \pm 10\) ms, which represent the time delay between consecutive bubbles reaching the rod interface (see Figure \ref{bubble_chain}a). Figures \ref{bubble_chain}a and b show the sequence of transport for a bubble series (with an equivalent diameter, \(D_b = 2.8\) mm) moving along a 3 mm diameter rod, showing the variation of capillary number (\(Ca\)) for the first (B1), second (B2), and third (B3) bubbles. The experimental results show a clear increase in velocity for the bubbles that come after, as indicated by the gradual rise in \(Ca\) from B1 to B3 (Figure \ref{bubble_chain}b). This acceleration is due to two simultaneous physical mechanisms: (i) Contact Line Lubrication and (ii) Wake-Induced Drag Reduction. When the leading bubble (B1) passes, it leaves behind small air volumes in the surface roughness of the rod. This gaseous phase modifies local wettability characteristics, reduces three-phase interfacial resistance, and establishes a low-friction slip boundary condition, thereby increasing the velocity of the secondary bubble. Moreover, the displacement of the surrounding fluid by the previous bubble produces a local hydrodynamic wake that features a negative pressure gradient and a region of fluid recirculation. Bubbles that enter this area of low pressure experience both a reduction in the forward viscous drag and an upward hydrodynamic lift, which causes their transport velocity along the rod surface to increase. To confirm the above explanation, a numerical study is presented in the following section.

\subsection{Numerical study on the bubble transport}

\begin{figure}[ht!]
\centering
\includegraphics[width=0.97\textwidth]{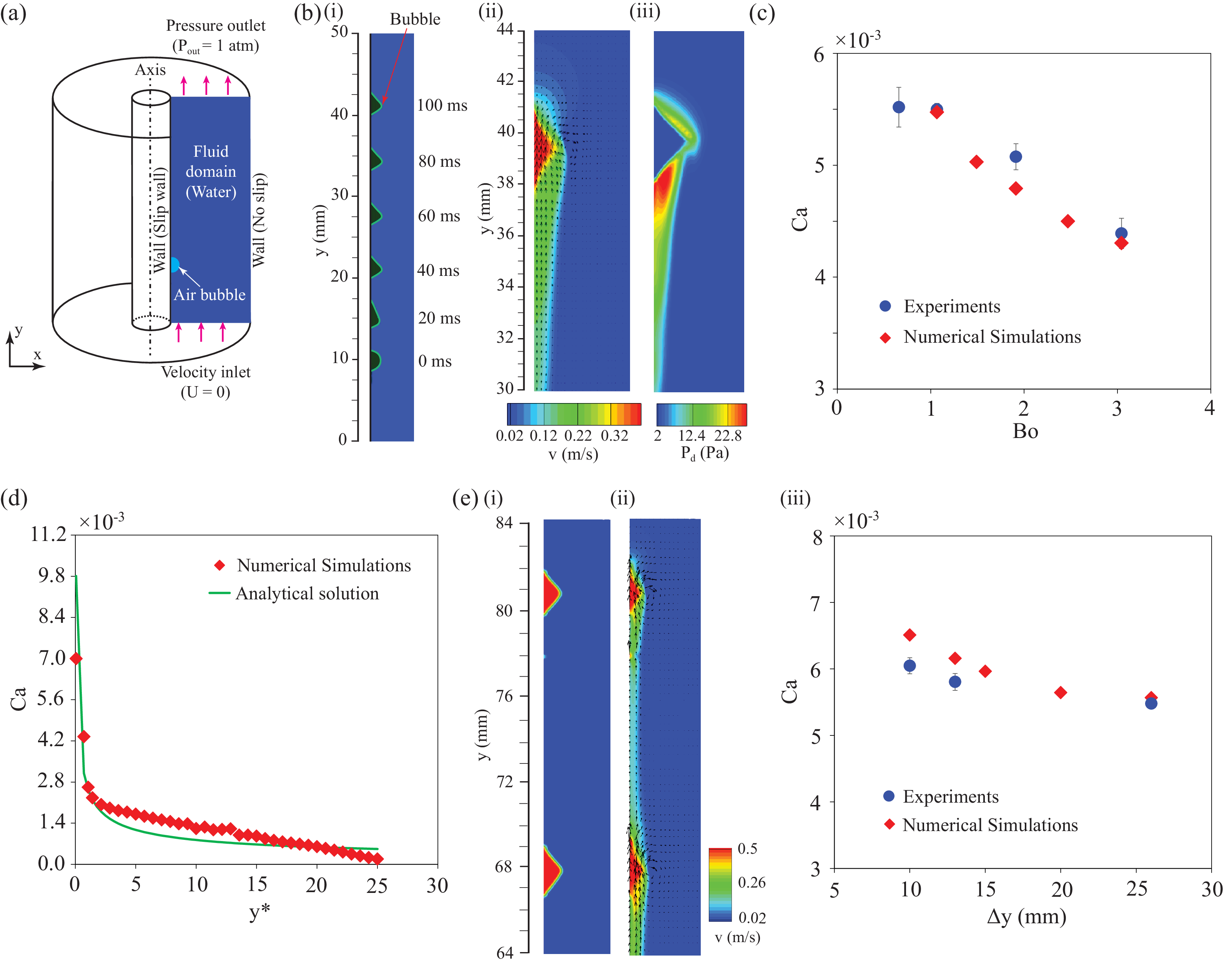}
\caption{Numerical studies and comparison with experimental results: (a) computational domain; (b) (i) bubble transport behavior, (ii) variation of velocity, (iii) variation of dynamic pressure; (c) comparison of the variation of capillary number as a function of Bond number from numerical simulations with experimental results; (d) variation of local capillary number behind the bubble from numerical simulations and analytical solutions ($y^* = \frac{y}{D_b}$); (e) (i) local variation of velocity across two consecutive bubbles, (ii) variation of the capillary number of the second bubble as a function of the initial axial distance between two consecutive bubbles for both experimental and numerical results.}
\label{fig:num-domain}
\end{figure}

A 2D axisymmetric numerical analysis was performed to understand the interaction of subsequent bubbles along the rod. The numerical methods have been discussed in the appendix (see S3). It has been taken as an axis-symmetric problem \cite{farhangi2010numerical}, assuming that the bubble wrap completely surrounds the rod. Figure \ref{fig:num-domain}a shows the computational domain used for the analysis. The width and height of the domain are $25$ mm and $100$ mm, respectively. The left-side wall (at $x = 1.5$ mm, as the $3$ mm rod has been considered for the numerical study) could be superhydrophobic, with a slip boundary condition (no shear force, and $\theta = 160$\textdegree). A zero inlet velocity boundary condition has been imposed on the bottom wall. The upper wall (at $y = 100$ mm) is defined as an outlet with atmospheric pressure \cite{chowdhury2022wettability}. Water and air were considered as the primary and secondary phases, respectively.

A semi-circular air bubble is initialized on the wall (at $x = 1.5$ mm), with a volume equal to that of the bubble in the experiment. The simulation has been performed for various $Bo$, and the corresponding $Ca$ has been compared with the experimental data, shown in Fig. \ref{fig:num-domain}c. Though the trends are similar, there are marginal differences in $Ca$ between the experimental data and the numerical studies. In the experiment, the plastron layer plays a significant role, whereas in the numerical simulation, it is not accurately represented. This is the reason for the deviation. Nevertheless, the numerical study does provide valuable insight. Figure~\ref{fig:num-domain} presents the simulated bubble shape, velocity field, and dynamic pressure distribution during transport. The simulations show that a recirculating wake (Fig.~\ref{fig:num-domain}b-ii) forms behind the moving bubble. The liquid velocity (Fig.~\ref{fig:num-domain}b-ii) and dynamic pressure (Fig.~\ref{fig:num-domain}b-iii) decrease along the direction behind the bubble, creating an upward flow field that benefits the motion of the next bubble. To better understand the wake-induced flow behavior, the variation of the local $Ca$ (velocity) behind the bubble using a scaling approach based on the diffusion equation \citep{ali2024diffusion}.

\begin{equation}
    \frac{\partial u}{\partial t} = \gamma \frac{\partial^2u}{\partial y^2}
    \label{diff}
\end{equation}

As the above equation is scaled as $\frac{\partial^2u}{\partial y^2}\sim \frac{u}{b^2}$ and $\frac{\partial u}{\partial t} \sim \frac{u}{t}$. Therefore, equation~\ref{diff} becomes $b \sim \sqrt{\gamma t}$. 

Where $ y$, $ u$, $b$, $t$, and $\gamma$ are the axial distance, the local velocity, the characteristic length scale of velocity variation, the time scale, and effective diffusivity, respectively. The characteristic time scale is converted as $t = \frac{y}{U}$ and $b\sim\sqrt{\frac{\gamma y}{U}}$, where $U$ is the velocity of the bubble. At terminal velocity, a bubble moves at a constant velocity. therefore, $b \sim \sqrt{y}$. Besides, the drag equation can be scaled as $F_D \sim \rho U u(y) b(y)$. At terminal velocity of the bubble $\rho$, $F_D$ and $U$ are constants, the drag equation becomes $u(y) b(y) = C$. Therefore, the variation in velocity in the $y$ direction behind the bubble can be expressed as follows (Eq.~\ref{local_u}).

\begin{equation}
    u(y) = \frac{C}{\sqrt{y}} 
    \label{local_u}
\end{equation}

Figure~\ref{fig:num-domain}d shows the variation in the local capillary number ($Ca = \frac{\mu u(y)}{\sigma}$), where the trend of the simulation data matches the analytical results.   
Figure ~\ref{fig:num-domain}e shows the contour of two consecutive bubbles transported on the superhydrophobic rod and the variation of $y$-velocity across two consecutive bubbles with the streamline. The wake and liquid circulation induced by the preceding bubble extend to the subsequent bubble (Fig.~\ref{fig:num-domain}e-ii), thus increasing the velocity of the subsequent bubble. However, the influence of the preceding bubble reduces as the distance between the two bubbles increases (Fig.~\ref{fig:num-domain}e-iii), as the effect of the bubble motion decreases over the distance from the bubble. It strengthens the earlier discussion of the higher velocity of the subsequent relative to the preceding bubble.

\section{Conclusion}
\label{sec:Con_ion}

In the present study, the transport and detachment behavior of air bubbles along submerged superhydrophobic cylindrical rods is investigated using experiments, analytical modeling, and numerical simulations. Since superhydrophobic surfaces can reduce interfacial pinning and promote bubble transportability \cite{li2019bubble,feng2020spontaneous}, we proposed that the curvature and dimension of the cylindrical substrate can influence bubble–surface interactions to hydrodynamic resistance, which controls bubble transport and detachment. We also hypothesized that the motion of a preceding bubble could affect the transport of a subsequent bubble in the surrounding liquid. The results confirm both hypotheses and show that bubble transport on cylindrical superhydrophobic surfaces is determined by coupled interfacial, geometric, and hydrodynamic effects, not merely by buoyancy.

Our study presents new insights into three aspects closely related to the transport of bubbles along cylindrical rods. Firstly, although earlier research has shown that superhydrophobicity enhances bubble mobility \cite{li2019bubble,feng2020spontaneous}, our findings indicate that the bubble-to-rod size ratio offers an additional geometric control over the transport process. Since the buoyancy force scales with the cube of the bubble diameter $(D_b^3)$, it also leads to greater bubble deformation and interfacial confinement, and increases the characteristic contact-line length. As a result, larger bubbles do not always move faster along the rod. The dependence of $Ca$ on $Bo$ also varies with the diameter of the rod. These results prove that bubble transport cannot be explained by Bo alone and instead arises from the interaction between the gravitational force and the geometry-dependent interfacial and hydrodynamic resistance.

Secondly, a force-balance framework was developed, which relates the velocity of bubble transport to buoyancy, hydrodynamic drag, and capillary/ contact-line hysteresis. The model reproduces the primary experimental trends and predicts the measured $Ca$ with deviations of less than 10\% for most conditions. More significantly, the formulation provides a physical basis for understanding the effects of rod diameter and bubble size, rather than treating the transport velocity as an empirical value. The detachment experiments also indicate that an increase in $Bo$ leads to detachment, while an increase in the characteristic rod size results in bubble retention. A regime map based on $Bo$ and the normalized rod diameter provides a simple way to represent the transition from stable transport to bubble detachment.

Next, we have shown that bubble transport becomes dynamically coupled when bubbles are generated one after another. Although the wakes behind freely rising bubbles have been well documented \cite{chai2015wake}, the effects of these wakes on the transport of bubbles in sequence along a curved superhydrophobic interface have not been studied in detail. Our results indicate that the bubble produces a recirculating flow field that extends downstream and alters the hydrodynamic environment the next bubble encounters. The following bubble speeds up and has a higher $Ca$ number than the preceding bubble; however, the magnitude of this increase in speed is inversely proportional to the distance between the bubbles. This experimental finding is reproduced by numerical simulations that resolve the wake and the associated velocity and pressure fields. The results show that the transport of a bubble on a superhydrophobic cylindrical surface is not an isolated event and that the motion of one bubble can influence the transport state of another through hydrodynamic coupling. Furthermore, the passage of the leading bubble may locally alter the interfacial state due to residual gas trapped in the surface texture, potentially contributing to the observed acceleration.

Future research can be directed toward bubble transport over arrays of cylindrical structures and more complex surface shapes to better reflect practical engineering systems. Extending the study by varying fluid properties, introducing turbulence or oscillatory flow conditions, and using a larger number of bubbles will improve our understanding of bubble behavior. In addition, studies of the effects of multiple-bubble interactions, surface durability, and the ability to scale up production will be crucial to applying laboratory findings to real-world situations. These efforts will help develop robust passive bubble-transport technologies for next-generation electrochemical energy systems \cite{kempler2024gas}, thermal management devices \cite{chen2015bubble}, microfluidic platforms \cite{zhao2022air}, and other gas-liquid transport applications.

\section*{Data Availability}

The data that support the findings of this study are available from the corresponding authors upon request.


\section*{Author Contributions}
\textbf{Rajalingam A:} Writing - original draft, Methodology, Investigation, Formal analysis, Data curation, Conceptualization, Software, Funding acquisition. \textbf{Sunghwan Jung:} Writing – review \& editing, Investigation, Formal analysis, Supervision, Conceptualization. \textbf{Pallab Sinha Mahapatra:} Writing – review \& editing, Investigation, Formal analysis, Supervision, Resources, Project administration, Funding acquisition.

\section*{Acknowledgment}
\label{sec:Ac_know}
The authors acknowledge the Anusandhan National Research Foundation (ANRF) for supporting research with funding for the National Postdoctoral Fellowship [Sanction. No: $PDF/2025/004895$]. P.S.M acknowledges the V. Ganesan Faculty Fellowship received from IIT Madras.


\section*{Appendix A. Supplementary Data} Experimental process validation and Numerical modeling details.(PDF)

\medskip


\end{document}